\documentclass[two column, prA]{revtex4}%
\usepackage{amsmath}
\usepackage{graphicx}
\usepackage{amsfonts}
\usepackage{amssymb}
\usepackage{color}%
\providecommand{\U}[1]{\protect\rule{.1in}{.1in}}
\ifx\pdfoutput\relax\let\pdfoutput=\undefined\fi
\newcount\msipdfoutput
\ifx\pdfoutput\undefined\else
\ifcase\pdfoutput\else
\ifx\paperwidth\undefined\else
\ifdim\paperheight=0pt\relax\else\pdfpageheight\paperheight\fi
\ifdim\paperwidth=0pt\relax\else\pdfpagewidth\paperwidth\fi
\fi\fi\fi
\begin{document}
\title{Photon position observable}
\author{Margaret Hawton}
\affiliation{Department of Physics, Lakehead University, Thunder Bay, ON, Canada, P7B 5E1}
\email{margaret.hawton@lakeheadu.ca}
\author{Vincent Debierre}
\affiliation{Department of Physics, Missouri University of Science and Technology, Rolla,
MO 65409-0640, USA}
\email{debierrev@mst.edu}

\begin{abstract}
In biorthogonal quantum mechanics, the eigenvectors of a quasi-Hermitian
operator and those of its adjoint are biorthogonal and complete and the
probability for a transition from a quantum state to any one of these
eigenvectors is positive definite$.$ We apply this formalism to the long
standing problem of the position observable in quantum field theory. The dual
bases are positive and negative frequency one-particle states created by the
field operator and its conjugate and biorthogonality is a consequence of their
commutation relations. In these biorthogonal bases the position operator is
covariant and the Klein-Gordon wave function is localized. We find that the
invariant probability for a transition from a one-photon state to a position
eigenvector is the first order Glauber correlation function, bridging the gap
between photon counting and the sensitivity of light detectors to
electromagnetic energy density.

\end{abstract}
\maketitle

\section{Introduction}

\label{sec:Intro}

Many applications and tests of quantum mechanics (QM) involve photons and some
require a basis of photon position eigenvectors \cite{Lundeen}. In spite of
its potential for direct application to experiment, it has been concluded that
there is no position observable or completeness relation for photons
\cite{BKP,ScullyZubairy}. In the assumed absence of photon number density,
energy density is used to define the photon wave function as the positive
frequency part of the Riemann-Silberstein vector $\mathbf{F}\propto
\mathbf{E}\pm\mathrm{i}c\mathbf{B}$ \cite{BB94,Sipe,BB96,SmithRaymer}. But a
consistent one-particle theory and position observable does exist for
Klein-Gordon (KG) particles \cite{Gitman,Mostafazadeh1}. We will show here
that this solution to the relativistic position observable problem can be
extended to photons.

The KG field is sometimes used as a simple model of a photon, for example in
curved space \cite{BirrellDavies}. The zeroth component of the conventional KG
four-current density that equals the difference between its particle and
antiparticle parts leads to an indefinite scalar product unless the KG field
is limited to positive frequencies. The positive frequency part of the
four-current density is sometimes interpreted as particle number density but
it can still be negative if components with two or more different frequencies
are added \cite{Holland,Colin,Nikolic}. However, if the KG scalar product is
recognized as a pseudo-scalar product within the framework of
pseudo{-Hermitian} QM \cite{Mostafazadeh1}, a positive definite particle
density \emph{is} obtained. This can be extended to an arbitrary linear
combination of positive and negative frequency terms if a new scalar product
is defined in which the particle and antiparticle contributions are added
\cite{Gitman,Mostafazadeh1,HalliwellOrtiz,HawtonPhysicaScripta}. Moreover,
Hegerfeldt's theorem tells us that restriction to positive frequencies leads
to instantaneous spreading of a particle's wave function \cite{Hegerfeldt}, so
negative frequencies are needed to obtain a causally evolving particle
density. A photon, like a neutral KG particle, is its own antiparticle so
these advances in our understanding of KG particle density are very relevant
to our goal here, which is to obtain analogous results for photons.

The KG wave function derived in \cite{Mostafazadeh1} is the projection of a
particle's state vector onto the eigenvectors of the Newton Wigner (NW)
position operator \cite{NewtonWigner}. NW found a position operator for KG
particles, but they concluded that the only photon position operator is the
Pryce operator whose vector components do not commute \cite{Pryce}, making the
simultaneous determination of photon position in all three directions of space
impossible. They had assumed spherically symmetrical position eigenstates for
photons, while photon position eigenvectors have an axis of symmetry like
twisted light \cite{Twisted}. Following the NW method with omission of the
spherical symmetry axiom, a photon position operator with commuting components
and eigenvectors that are cylindrically symmetrical in $\mathbf{k}${-space}
can be constructed \cite{HawtonPosOp}. Since spin and orbital angular momentum
are not separately observable \cite{CohenQED1}, its eigenvectors have only
definite total angular momentum along some fixed but arbitrary axis
\cite{HawtonBaylis}.

Here all calculations will be performed in the physical Hilbert space of
solutions to the wave equation using a scalar product of the form derived in
\cite{Mostafazadeh1} and biorthogonal QM \cite{Brody} summarized here as
follows: The eigenvectors of a quasi-Hermitian \cite{quasi} operator $\hat{O}$
and its adjoint $\hat{O}^{\dagger}$ are not orthogonal, as is the case for
conventional Hermitian operators, but biorthogonal. This means that, given the
eigenvector equations%
\begin{align}
\hat{O}|\omega_{i}\rangle &  =\omega_{i}|\omega_{i}\rangle,\\
\hat{O}^{\dagger}|\tilde{\omega}_{j}\rangle &  =\omega_{j}|\tilde{\omega}%
_{j}\rangle
\end{align}
we have $\langle\tilde{\omega}_{j}|\omega_{i}\rangle=\delta_{ji}\langle
\tilde{\omega}_{i}|\omega_{i}\rangle$ and the completeness relation
$\widehat{1}=\sum_{i}\left\vert \omega_{i}\right\rangle \left\langle
\widetilde{\omega}_{i}\right\vert /\langle\tilde{\omega}_{i}|\omega_{i}%
\rangle.$ An arbitrary state $\left\vert \psi\right\rangle $ has an associated
state $\left\vert \widetilde{\psi}\right\rangle $. If an arbitrary state
vector is expanded as $\left\vert \psi\right\rangle =\sum_{i}c_{i}\left\vert
\omega_{i}\right\rangle $ in the Hilbert space $\mathcal{H}$ then in
biorthogonal QM its associated state is $\left\vert \widetilde{\psi
}\right\rangle =\sum_{i}c_{i}\left\vert \widetilde{\omega}_{i}\right\rangle
\in\mathcal{H}^{\ast}$ where $c_{i}=\left\langle \widetilde{\omega}_{i}%
|\psi\right\rangle =\left\langle \omega_{i}|\widetilde{\psi}\right\rangle .$
Using these expansions it is straightforward to verify that $\left\langle
\widetilde{\psi}_{1}|\psi_{2}\right\rangle =\left\langle \psi_{1}%
|\widetilde{\psi}_{2}\right\rangle .$ The probability for a transition from a
quantum state $\left\vert \psi\right\rangle $ to an eigenvector $\left\vert
\tilde{\omega}_{i}\right\rangle ${ of }$\hat{O}^{\dagger}$ is
\begin{equation}
p_{i}=\frac{\left\vert \left\langle \widetilde{\omega}_{i}|\psi\right\rangle
\right\vert ^{2}}{\left\langle \widetilde{\psi}|\psi\right\rangle \left\langle
\widetilde{\omega}_{i}|\omega_{i}\right\rangle }. \label{p}%
\end{equation}
A generic operator can be written in the form
\begin{equation}
\widehat{F}=\sum_{i,j}f_{ij}\left\vert \omega_{i}\right\rangle \left\langle
\widetilde{\omega}_{j}\right\vert \label{F}%
\end{equation}
where $f_{ij}$ can be viewed as a matrix \cite{Brody}. In Section
\ref{sec:Scalar} we will apply this formalism to the biorthogonal position
eigenvectors $\left\vert \phi\left(  x\right)  \right\rangle =\widehat{\phi
}\left(  x\right)  |0\rangle$ and $\left\vert \widetilde{\phi}\left(
x\right)  \right\rangle =\left\vert \pi\left(  x\right)  \right\rangle
=\widehat{\pi}\left(  x\right)  |0\rangle$ where $x^{\mu}=\left(
ct,\mathbf{x}\right)  ,$ $\widehat{\phi}\left(  x\right)  $ is a field
operator, $\widehat{\pi}\left(  x\right)  $ is its conjugate momentum
operator, and $|0\rangle$ is the vacuum state.

{The rest of this paper is organized as follows: In Section \ref{sec:KG} }KG
wave mechanics,{ with the field rescaled here to facilitate application to
particles with zero mass, is reviewed. In Section~\ref{sec:Scalar}}
biorthogonality of the one-particle states created by the field operator and
its conjugate momentum are examined and the covariant position operator and
positive definite probability density are derived. A s{econd quantized
formalism is used to facilitate future application to multiparticle problems
such as entanglement. In Section~\ref{sec:Photons} the KG position observable
discussed in Sections \ref{sec:KG} and \ref{sec:Scalar} is extended to
photons. In Section~\ref{sec:Emission} the wave function of the photon emitted
by an atom is discussed and in Section~\ref{sec:Ccl} we conclude.}

The configuration space scalar field and four-potential will be called
$\phi\left(  x\right)  $ and $A^{\mu}\left(  x\right)  $. State vectors such
as $\left\vert \psi\left(  t\right)  \right\rangle $ and position eigenvectors
such as $\left\vert \phi\left(  x\right)  \right\rangle $ and $\left\vert
\mathbf{E}_{\lambda}\left(  x\right)  \right\rangle $ introduced in Sections
\ref{sec:Scalar} and \ref{sec:Photons} are given as expansions in Fourier
space. The function $\psi\left(  x\right)  $ used in
\cite{FeshbachVillars,Mostafazadeh1} to represent the scalar field will be
reserved for the wave function that equals the projection of a particle's
state vector onto a basis of position eigenvectors. The KG and photon wave
functions, $\psi\left(  x\right)  $ and $\mathbf{\psi}_{\perp}\left(
x\right)  $ respectively, are proportional to probability amplitudes, with
units that differ from those of $\phi\left(  x\right)  $, $A^{\mu}\left(
x\right)  $ and their spacetime derivatives.

\section{Klein-Gordon wave mechanics}

\label{sec:KG}

We will start with a review of the KG position observable problem. The KG
equation%
\begin{equation}
\partial_{\mu}\partial^{\mu}\phi\left(  x\right)  +\frac{m^{2}c^{2}}{\hbar
^{2}}\phi\left(  x\right)  =0 \label{KGeq}%
\end{equation}
describes charged and neutral particles with zero spin (pions). Here covariant
notation and the mostly minus convention are used in which $x^{\mu}=x=\left(
ct,\mathbf{x}\right)  ,$ $\partial_{\mu}=\left(  \partial_{ct},\mathbf{\nabla
}\right)  ,$ $m$ is the mass of the KG particle, $c$ is the speed of light,
$2\pi\hbar$ is Planck's constant and $f_{1}\overleftrightarrow{\partial}_{\mu
}f_{2}\equiv f_{1}\left(  \partial_{\mu}f_{2}\right)  -\left(  \partial_{\mu
}f_{1}\right)  f_{2}.$ The function $\phi\left(  x\right)  $ is any scalar
field that satisfies the KG equation (\ref{KGeq}).\textbf{ }The four-density
\begin{equation}
J_{KG}^{\mu}\left(  x\right)  =\mathrm{i}g\phi\left(  x\right)  ^{\ast
}\overleftrightarrow{\partial}^{\mu}\phi\left(  x\right)  , \label{KG}%
\end{equation}
satisfies a continuity equation. Plane wave normal mode solutions to
(\ref{KGeq}) proportional to $\exp\left(  -\mathrm{i}\omega t\right)  $ are
referred to as positive frequency solutions, while those proportional to
$\exp\left(  \mathrm{i}\omega t\right)  $ are negative frequency. Completeness
requires that both positive and negative frequency modes be included. Their
contributions to $J_{KG}^{0}\left(  x\right)  $ are of opposite sign, so
$J_{KG}^{0}\left(  x\right)  $ is interpreted as charge density and the
quantity $g$ in (\ref{KG}) is set equal to $qc/\hbar$ for particles of charge
$q$.

If only particles, {as opposed to both particles and antiparticles}, are to be
considered, then the KG field can be restricted to positive frequencies and
the scalar product \cite{BirrellDavies}%
\begin{equation}
\left(  \phi_{1},\phi_{2}\right)  _{KG}=\frac{\mathrm{i}}{\hbar}\int%
_{t}\mathrm{d}\mathbf{x}\,\phi_{1}\left(  x\right)  ^{\ast}%
\overleftrightarrow{\partial}_{t}\phi_{2}\left(  x\right)  \label{KGsp}%
\end{equation}
is positive definite. Here $t$ denotes a spacelike hyperplane of simultaneity
at instant $t.$ The integrand of (\ref{KGsp}) looks like a particle density
but this is misleading since, as noted in the Introduction, $J_{KG}^{0}\left(
x\right)  $ is not positive definite for positive frequency fields.

The problem of a probability interpretation for KG particles has a long
history. Lack of a probability interpretation led Dirac to derive his
celebrated equation for spin half particles, but this does not solve the
problem for KG fields. In a seminal paper intended to clarify the confusion
about relativistic wave mechanics, Feshbach and Villars reviewed the two
component formalism that separates the wave function into its particle and
antiparticle parts for charged or for neutral particles \cite{FeshbachVillars}%
. Since then various strategies have been employed to derive a positive
definite probability density. The four-current density can be redefined so
that its zeroth component is positive definite \cite{GGP}, but this
construction has no apparent physical basis and it fails if $m=0$
\cite{HenningWolf}. It has been proposed that for charged pions only positive
definite eigenstates of the Hamiltonian are physical \cite{SUL}. A new
$J^{\mu}\left(  x\right)  $ was derived that does not require separation of
the field into positive and negative frequency parts \cite{Mostafazadeh2} so
it can be applied to the real fields that describe neutral pions. If
$\phi\left(  x\right)  $ is restricted to positive frequencies it reduces to
(\ref{KG}) so this $J^{\mu}\left(  x\right)  $ that describes an arbitrary
linear combination of positive and negative frequency fields, including real
fields, will be used here.

Working in the two component formalism with a pseudo-Hermitian Hamiltonian
Mostafazadeh \cite{Mostafazadeh1} defined the positive-definite Hermitian
operator
\begin{equation}
\widehat{D}\equiv-\nabla^{2}+m^{2}c^{2}/\hslash^{2} \label{D}%
\end{equation}
in terms of which the KG equation is $\left(  \widehat{D}+\partial_{ct}%
^{2}\right)  \psi=0.$ He derived the conjugate field
\begin{equation}
\phi_{c}\left(  x\right)  =\mathrm{i}\widehat{D}^{-1/2}\partial_{ct}%
\phi\left(  x\right)  \label{psi_c}%
\end{equation}
such that if $\phi=\phi^{+}+\phi^{-}$ then $\phi_{c}=\phi^{+}-\phi^{-}.$ This
implies that $\phi_{c}$ is a scalar. It is then straightforward to verify that
$\phi_{c}$ satisfies the KG equation and that%
\begin{equation}
J^{\mu}\left(  x\right)  =\frac{\mathrm{i}}{\hbar}\phi\left(  x\right)
^{\ast}\overleftrightarrow{\partial}^{\mu}\phi_{c}\left(  x\right)  \label{M}%
\end{equation}
satisfies the continuity equation $\partial_{\mu}J^{\mu}=0$. Up to a constant
that just scales $J^{\mu},$ (\ref{M}) is the expression derived in
\cite{Mostafazadeh2}. Like (\ref{KG}) $J^{\mu}\left(  x\right)  $ is
manifestly covariant. It was proved in \cite{Mostafazadeh1,Mostafazadeh2} that
the scalar product
\begin{equation}
\left(  \phi_{1},\phi_{2}\right)  =\frac{\mathrm{i}}{\hbar}\int_{t}%
\mathrm{d}\mathbf{x}\,\phi_{1}\left(  x\right)  ^{\ast}%
\overleftrightarrow{\partial}_{t}\phi_{2c}\left(  x\right)  \label{Msp}%
\end{equation}
is positive definite, time independent and can be written in covariant form
as
\begin{equation}
\left(  \phi_{1},\phi_{2}\right)  =\frac{\mathrm{i}}{\hbar}\int_{n}%
{\mathrm{d}}\sigma n_{\mu}\phi_{1}\left(  x\right)  ^{\ast}%
\overleftrightarrow{\partial}^{\mu}\phi_{2c}\left(  x\right)
\label{invariant_sp}%
\end{equation}
where $n$ is an arbitrary spacelike hyperplane with normal $n^{\mu}$, in other
words, it is a Cauchy surface. The infinitesimal volume elements
{${\mathrm{d}}\sigma\equiv\mathrm{d}x$} are invariant. When restricted to
positive frequencies (\ref{M}) reduces to (\ref{KG}) and (\ref{Msp}) reduces
to (\ref{KGsp}). Using (\ref{KGeq}), (\ref{psi_c}) and $\phi_{c}=\phi^{+}%
-\phi^{-}$, the scalar product (\ref{Msp}) can be written as
\begin{equation}
\left(  \phi_{1},\phi_{2}\right)  =\frac{2c}{\hbar}\sum_{\epsilon=\pm
}\left\langle \phi_{1}^{\epsilon}|\widehat{D}^{1/2}\phi_{2}^{\epsilon
}\right\rangle \label{sp}%
\end{equation}
where%
\begin{equation}
\left\langle \chi_{1}|\chi_{2}\right\rangle =\int\mathrm{d}\mathbf{x}%
\,\chi_{1}^{\ast}\left(  \mathbf{x}\right)  \chi_{2}\left(  \mathbf{x}\right)
. \label{nonrel_sp}%
\end{equation}

The non-relativistic Hilbert space is the vector space of square integrable
continuous functions with the scalar product (\ref{nonrel_sp}). In the
relativistic Hilbert space the scalar product used here is (\ref{sp}). These
scalar products can be evaluated in configuration space or in $\mathbf{k}%
$-space.\textbf{ }The covariant Fourier transform is%
\begin{equation}
\phi^{\epsilon}\left(  x\right)  =\int\frac{\mathrm{d}\mathbf{k}\,}{\left(
2\pi\right)  ^{3}2\omega_{\mathbf{k}}}\pi^{\epsilon}\left(  \mathbf{k}\right)
\mathrm{e}^{-\mathrm{i}\epsilon\left(  \omega_{\mathbf{k}}t-\mathbf{k\cdot
x}\right)  }. \label{FT}%
\end{equation}
Since $\phi^{\epsilon}\left(  x\right)  $ is a scalar and $\mathrm{d}%
\mathbf{k}\,/\left[  \left(  2\pi\right)  ^{3}2\omega_{\mathbf{k}}\right]  $
is invariant, $\pi^{\epsilon}\left(  \mathbf{k}\right)  \in\mathcal{H}^{\ast}$
is a scalar, analogous to the transformation properties of photons
\cite{HawtonLorentzInv08}. For $\omega_{\mathbf{k}}=\sqrt{\mathbf{k}^{2}%
c^{2}+m^{2}c^{4}/\hbar^{2}}$ the function $\phi^{\epsilon}\left(  x\right)  $
satisfies the KG equation and (\ref{sp}) can be written as
\begin{equation}
\left(  \phi_{1},\phi_{2}\right)  =\frac{1}{\hbar}\sum_{\epsilon=\pm}\int%
\frac{\mathrm{d}\mathbf{k}\,}{\left(  2\pi\right)  ^{3}2\omega_{\mathbf{k}}%
}\pi_{1}^{\epsilon\ast}\left(  \mathbf{k}\right)  \pi_{2}^{\epsilon}\left(
\mathbf{k}\right)  . \label{k_sp}%
\end{equation}
In the biorthogonal formalism the bases $\left\{  \phi_{\mathbf{x}_{j}%
}^{\epsilon}\left(  \mathbf{k}\right)  \right\}  =\left\{  \pi_{\mathbf{x}%
_{j}}^{\epsilon}\left(  \mathbf{k}\right)  /\omega_{\mathbf{k}}\right\}
\in\mathcal{H}$ and $\left\{  \pi_{\mathbf{x}_{j}}^{\epsilon}\left(
\mathbf{k}\right)  \right\}  \in\mathcal{H}^{\ast}$ are biorthogonal and
complete and the Hermitian adjoint of an operator is its complex conjugate
transpose \cite{Brody}. Since the scalar product (\ref{sp}) is positive
definite, standard QM can be recovered if a nontrivial metric operator\textbf{
}$\left\langle .|\Theta.\right\rangle $\textbf{ }is introduced
\cite{GeyerScholtz}. In this metric formulation the basis is $\left\{
\pi_{\mathbf{x}_{j}}^{\epsilon}\left(  \mathbf{k}\right)  \right\}  $ and
operators are Hermitian. With the flat metric\textbf{ }$\left\langle
.|.\right\rangle $\textbf{ }operators representing observables can be
non-Hermitian with biorthogonal eigenvectors. The norm and orthogonality of
the elements of the Hilbert space\ and the concept of Hermiticity are
determined by the definition of scalar product. Newton and Wigner defined the
KG fields $\pi_{\mathbf{x}_{j}}^{\epsilon}\left(  \mathbf{k}\right)  \propto
e^{-\mathrm{i}\mathbf{k\cdot x}_{j}}\omega_{\mathbf{k}}^{1/2}$ that satisfy
$\left(  \phi_{1},\phi_{2}\right)  \propto\delta\left(  \mathbf{x}%
_{1}-\mathbf{x}_{2}\right)  \ $and are eigenvectors of the position operator
$\mathrm{i}\nabla_{\mathbf{k}}-\frac{\mathrm{i}c^{2}}{2\omega_{\mathbf{k}}%
^{2}}\mathbf{k}$ \cite{NewtonWigner}.\textbf{ }Hermiticity of the NW position
operator with eigenvectors of this form is discussed by Pike and Sarkar
\cite{PikeSarkar}. The NW position operator has played a central role in the
discussion relativistic particle position since its publication in 1949
\cite{NewtonWigner}, but this operator is not covariant and its eigenvectors
are not localized. We will show in the next section that the covariant
commutation relations are consistent with (\ref{sp}) and that the formalism of
biorthogonal QM leads to a covariant position operator. A second quantized
version of the biorthogonal formulation in \cite{Brody} will be used to
facilitate applications in quantum optics and understanding of the
relationship of the classical wave equation to quantum field theory (QFT).

\section{KG position eigenvectors}

\label{sec:Scalar} In QFT particles are created at a point in spacetime by a
field operator or its canonical conjugate. The interaction picture (IP) scalar
field operators $\widehat{\phi}\left(  x\right)  $ and $\widehat{\pi}\left(
x\right)  =\partial_{t}\widehat{\phi}\left(  x\right)  $ will be written as
\begin{subequations}
\label{eq:phipi}%
\begin{align}
\widehat{\phi}\left(  x\right)   &  =\sqrt{\hbar}\int\frac{\mathrm{d}%
\mathbf{k}}{\left(  2\pi\right)  ^{3}2\omega_{\mathbf{k}}}\mathrm{e}%
^{\mathrm{i}\left(  \omega_{\mathbf{k}}t-\mathbf{k}\cdot\mathbf{x}\right)
}\widehat{a}^{\dagger}\left(  \mathbf{k}\right)  +\mathrm{H.c.},\label{phi}\\
\widehat{\pi}\left(  x\right)   &  =\mathrm{i}\sqrt{\hbar}\int\frac
{\mathrm{d}\mathbf{k}}{\left(  2\pi\right)  ^{3}2}\mathrm{e}^{\mathrm{i}%
\left(  \omega_{\mathbf{k}}t-\mathbf{k}\cdot\mathbf{x}\right)  }%
\widehat{a}^{\dagger}\left(  \mathbf{k}\right)  +\mathrm{H.c.}%
\end{align}
where $\mathrm{H.c.}$ is the Hermitian conjugate\textbf{ }and the covariant
normalization condition is \cite{ItzyksonZuber}%
\end{subequations}
\begin{equation}
\left[  \widehat{a}\left(  \mathbf{k}\right)  ,\widehat{a}^{\dagger}\left(
\mathbf{q}\right)  \right]  =\left(  2\pi\right)  ^{3}2\omega_{\mathbf{k}%
}\,\delta\left(  \mathbf{k}-\mathbf{q}\right)  . \label{k_communation}%
\end{equation}
On the $t$ hyperplane the field operators satisfy the commutation relations
\begin{equation}
\left[  \widehat{\phi}\left(  x\right)  ,\widehat{\pi}\left(  y\right)
\right]  =\mathrm{i}\hbar\delta\left(  \mathbf{x}-\mathbf{y}\right)  .
\label{x_commutation}%
\end{equation}

If the vacuum state $\left\vert 0\right\rangle $ is defined by the condition
$\forall\mathbf{k}\widehat{a}\left(  \mathbf{k}\right)  \left\vert
0\right\rangle =0$ then the field operators create one-particle states in this
vacuum. In the IP the basis vectors are time dependent \cite{CohenQED1}. To
accomodate positive and negative frequency wavefunctions, $\epsilon=\pm$
states will be defined as
\begin{subequations}
\label{phipi}%
\begin{align}
\left\vert \phi^{\epsilon}\left(  x\right)  \right\rangle  &  =\sqrt{\hbar
}\int\frac{\mathrm{d}\mathbf{k}}{\left(  2\pi\right)  ^{3}2\omega_{\mathbf{k}%
}}\mathrm{e}^{\mathrm{i}\epsilon\left(  \omega_{\mathbf{k}}t-\mathbf{k}%
\cdot\mathbf{x}\right)  }\left\vert 1_{\mathbf{k}}\right\rangle
,\label{phi_epsilon}\\
\left\vert \pi^{\epsilon}\left(  x\right)  \right\rangle  &  =\sqrt{\hbar}%
\int\frac{\mathrm{d}\mathbf{k}}{\left(  2\pi\right)  ^{3}2}\mathrm{e}%
^{\mathrm{i}\epsilon\left(  \omega_{\mathbf{k}}t-\mathbf{k}\cdot
\mathbf{x}\right)  }\left\vert 1_{\mathbf{k}}\right\rangle ,
\label{pi_epsilon}%
\end{align}
where $\left\vert \phi^{+}\left(  x\right)  \right\rangle \equiv\widehat{\phi
}^{-}\left(  x\right)  \left\vert 0\right\rangle $ and $\left\vert
\pi^{\epsilon}\left(  x\right)  \right\rangle \equiv c\widehat{D}%
^{1/2}\left\vert \phi^{\epsilon}\left(  x\right)  \right\rangle $ so that the
phase factor $\mathrm{i}$ is absorbed into the bases and%
\end{subequations}
\begin{equation}
\mathrm{i}\partial_{t}\left\vert \pi^{\epsilon}\left(  x\right)  \right\rangle
=-\epsilon c\widehat{D}^{1/2}\left\vert \pi^{\epsilon}\left(  x\right)
\right\rangle . \label{t_dependence}%
\end{equation}
With these definitions $\left\langle \pi^{+}\left(  x\right)  |\psi
^{+}\right\rangle \ $is positive frequency while
\begin{equation}
\left\langle \pi^{-}\left(  x\right)  |\psi^{-}\right\rangle =\left\langle
\pi^{+}\left(  x\right)  |\psi^{+}\right\rangle ^{\ast}=\left\langle \psi
^{+}|\pi^{+}\left(  x\right)  \right\rangle \label{negative}%
\end{equation}
is negative frequency where $\epsilon=+$ refers to a particle arriving from
the past and absorbed on $n$, while $\epsilon=-$ refers to a particle emitted
on $n$ and propagating into the future.\textbf{ }These basis vectors are
biorthogonal in the sense that{ }%
\begin{equation}
\left\langle \pi^{\epsilon}\left(  x\right)  |\phi^{\epsilon^{\prime}}\left(
y\right)  \right\rangle =\frac{\hbar}{2}{\delta_{n}\left(  x-y\right)  }%
\delta_{\epsilon\epsilon^{\prime}}. \label{x_sp}%
\end{equation}
The notation ${\delta_{n}\left(  x-y\right)  }$ is defined to {select }$x$ and
$y$ {such} that $x^{\mu}=y^{\mu}$ on the hyperplane with normal $n_{\mu}.$
Since $\left\vert \phi^{\epsilon}\left(  x\right)  \right\rangle $ and
$\left\vert \pi^{\epsilon}\left(  x\right)  \right\rangle $ are biorthogonal,
they satisfy the completeness relation%
\begin{equation}
\widehat{1}=\frac{2}{\hbar}\sum_{\epsilon=\pm}\int\mathrm{d}\mathbf{x}%
\left\vert \phi^{\epsilon}\left(  x\right)  \right\rangle \left\langle
\pi^{\epsilon}\left(  x\right)  \right\vert \label{x_complete}%
\end{equation}
where the factor $2/\hbar$ is due to normalization (see (\ref{x_sp})).

There is a direct correspondence between the scalar product (\ref{sp}) and the
vacuum expectation value of the QFT commutator. Substitution of
(\ref{pi_epsilon}), (\ref{t_dependence}) and (\ref{negative}) in the vacuum
expectation value of (\ref{x_commutation}) gives
\begin{align}
\left\langle \phi\left(  x\right)  |\pi\left(  y\right)  \right\rangle  &
=\left\langle 0\left\vert \widehat{\phi}^{+}\left(  x\right)  \widehat{\pi
}^{-}\left(  y\right)  -\widehat{\pi}^{+}\left(  y\right)  \widehat{\phi}%
^{-}\left(  x\right)  \right\vert 0\right\rangle \nonumber\\
&  =\left\langle \phi^{+}\left(  x\right)  |\pi^{+}\left(  y\right)
\right\rangle {+}\left\langle \pi^{+}\left(  y\right)  |\phi^{+}\left(
x\right)  \right\rangle \nonumber\\
&  =\left\langle \phi^{+}\left(  x\right)  |\pi^{+}\left(  y\right)
\right\rangle {+}\left\langle \phi^{-}\left(  x\right)  |\pi^{-}\left(
y\right)  \right\rangle .\label{c}%
\end{align}
On the $t$ hyperplane events $x^{\mu}=\left(  ct_{x},\mathbf{x}\right)  $ and
$y^{\mu}=\left(  ct_{y},\mathbf{y}\right)  $ appear simultaneous but an
inertial observer with velocity c$\beta$ will see these events as time
ordered. Since the Fourier space integrand of $\left\langle \phi^{+}\left(
x\right)  |\pi^{+}\left(  y\right)  \right\rangle $ is proportional to
$e^{-\mathrm{i}\omega_{\mathbf{k}}\left(  t_{x}-t_{y}\right)  }$ while that of
$\left\langle \pi^{-}\left(  y\right)  |\phi^{-}\left(  x\right)
\right\rangle $ will be seen as proportional to $e^{\mathrm{i}\omega
_{\mathbf{k}}\left(  t_{x}-t_{y}\right)  },$ if $t_{x}>t_{y}$ the first term
is positive frequency $\left(  \epsilon=+\right)  $ while the second is
negative frequency $\left(  \epsilon=-\right)  $. This assignment is not
unique, since an observer with velocity $-c\beta$ will see the opposite time
order. Thus the covariant scalar product (\ref{c}) should be a sum over
$\epsilon=\pm$, consistent with (\ref{sp}) but inconsistent with (\ref{KGsp}).
Based on (\ref{c}) or (\ref{invariant_sp}) a particle is either emitted or
absorbed at $x$ on $n$ and there are no $\epsilon=+/\epsilon=-$ cross terms in
the scalar product.

It can be verified by substitution that the basis states (\ref{phipi}) are
eigenvectors of a position operator of the form (\ref{F}),%
\begin{equation}
\widehat{\mathbf{x}}=\frac{2}{\hbar}\sum_{\epsilon=\pm}\int\mathrm{d}%
\mathbf{x\ x}\left\vert \phi^{\epsilon}\left(  x\right)  \right\rangle
\left\langle \pi^{\epsilon}\left(  x\right)  \right\vert ,
\label{PositionOperator}%
\end{equation}
and its adjoint, that is
\begin{subequations}
\label{scalar_eigenvectors}%
\begin{align}
\widehat{\mathbf{x}}\left\vert \phi^{\epsilon}\left(  x\right)  \right\rangle
&  =\mathbf{x}\left\vert \phi^{\epsilon}\left(  x\right)  \right\rangle
,\label{eq:XEigen}\\
\widehat{\mathbf{x}}^{\dagger}\left\vert \pi^{\epsilon}\left(  x\right)
\right\rangle  &  =\mathbf{x}\left\vert \pi^{\epsilon}\left(  x\right)
\right\rangle , \label{eq:XDaggerEigen}%
\end{align}
consistent with their biorthogonality. Any one-particle state can therefore be
projected onto the position bases as
\end{subequations}
\begin{subequations}
\label{x_basis}%
\begin{align}
\left\vert \psi\left(  t\right)  \right\rangle  &  =\widehat{1}\left\vert
\psi\left(  t\right)  \right\rangle \nonumber\\
&  =\frac{2}{\hbar}\sum_{\epsilon=\pm}\int\mathrm{d}\mathbf{x}\left\vert
\phi^{\epsilon}\left(  x\right)  \right\rangle \left\langle \pi^{\epsilon
}\left(  x\right)  |\psi\left(  t\right)  \right\rangle ,\\
\left\vert \widetilde{\psi}\left(  t\right)  \right\rangle  &  =\widehat{1}%
\left\vert \widetilde{\psi}\left(  t\right)  \right\rangle \nonumber\\
&  =\frac{2}{\hbar}\sum_{\epsilon=\pm}\int\mathrm{d}\mathbf{x}\left\vert
\pi^{\epsilon}\left(  x\right)  \right\rangle \left\langle \phi^{\epsilon
}\left(  x\right)  |\widetilde{\psi}\left(  t\right)  \right\rangle .
\end{align}
The wave function
\end{subequations}
\begin{equation}
\psi^{\epsilon}\left(  x\right)  =\left\langle \pi^{\epsilon}\left(  x\right)
|\psi\left(  t\right)  \right\rangle \label{wf}%
\end{equation}
completely describes the one-photon state $\left\vert \psi\left(  t\right)
\right\rangle $ in the $\left\{  \left\vert \phi^{\epsilon}\left(  x\right)
\right\rangle \right\}  $ basis of position eigenvectors. It may have positive
frequency and negative frequency components. According to the rules of
biorthogonal QM outlined in Section \ref{sec:Intro} we have the equality
\begin{equation}
\left\langle \phi^{\epsilon}\left(  x\right)  |\widetilde{\psi}\left(
t\right)  \right\rangle =\left\langle \pi^{\epsilon}\left(  x\right)
|\psi\left(  t\right)  \right\rangle . \label{amplitude}%
\end{equation}
Using (\ref{x_sp}), (\ref{x_basis}) and (\ref{amplitude}) the squared norm of
$\left\vert \psi\left(  t\right)  \right\rangle ,$
\begin{equation}
\left\langle \psi|\widetilde{\psi}\right\rangle =\frac{2}{\hbar}\sum
_{\epsilon=\pm}\int\mathrm{d}\mathbf{x}\left\vert \left\langle \pi^{\epsilon
}\left(  x\right)  |\psi^{\epsilon}\left(  t\right)  \right\rangle \right\vert
^{2},
\end{equation}
and the probability density,%
\begin{equation}
p^{\epsilon}\left(  x\right)  =\frac{2}{\hbar\left\langle \psi|\widetilde{\psi
}\right\rangle }\left\vert \left\langle \pi^{\epsilon}\left(  x\right)
|\psi\left(  t\right)  \right\rangle \right\vert ^{2}, \label{p_KG}%
\end{equation}
are positive definite. It is the invariant $p\left(  x\right)  $ given by
(\ref{p_KG}), not the zeroth component of the four-current, $J^{0}\left(
x\right)  ,$ that describes particle density.

Since this application of biorthogonal QM is based on an invariant scalar,
product the QM that it describes is covariant. In particular, the wave
function of a plane wave, $\left\vert 1_{\mathbf{k}}\right\rangle ,$ is
$\left\langle 1_{\mathbf{k}}|1_{\mathbf{q}}\right\rangle =\left(  2\pi\right)
^{3}2\omega_{\mathbf{k}}\,\delta\left(  \mathbf{k}-\mathbf{q}\right)  \ $in
Fourier space and $\left\langle \phi\left(  x\right)  |1_{\mathbf{q}%
}\right\rangle =\sqrt{\hbar}\mathrm{e}^{-\mathrm{i}\left(  \omega_{\mathbf{k}%
}t-\mathbf{q}\cdot\mathbf{x}\right)  }$ in configuration space and a localized
state, $\left\vert \phi\left(  y\right)  \right\rangle ,$ is $\left\langle
1_{\mathbf{k}}|\phi\left(  y\right)  \right\rangle =\sqrt{\hbar}%
\mathrm{e}^{\mathrm{i}\left(  \omega_{\mathbf{k}}t-\mathbf{k}\cdot
\mathbf{y}\right)  }$ in Fourier space and $\left\langle \pi\left(  x\right)
|\phi\left(  y\right)  \right\rangle =\frac{\hbar}{2}{\delta_{n}\left(
x-y\right)  }$ in configuration space. Eqs. (\ref{x_complete}) and
(\ref{PositionOperator}) can be generalized to%
\begin{align}
\widehat{1}  &  =\frac{2}{\hbar}\int\mathrm{d}\sigma\left\vert \phi\left(
x\right)  \right\rangle \left\langle -\mathrm{i}\epsilon n_{\mu}\partial^{\mu
}\phi\left(  x\right)  \right\vert ,\label{invariant_1}\\
\widehat{x}_{i}  &  =\frac{2}{\hbar}\int\mathrm{d}\sigma x_{i}\left\vert
\phi\left(  x\right)  \right\rangle \left\langle -\mathrm{i}\epsilon n_{\mu
}\partial^{\mu}\phi\left(  x\right)  \right\vert \label{invariant_x}%
\end{align}
respectively where $x_{\mu}n^{\mu}=ct_{0}$ on the hyperplane with normal
$n_{\mu}$ and $t_{0}$ is the hyperplane of simultaneity
\cite{Fleming,HawtonPhysicaScripta}. The matrix representing the position
observable in configuration space is
\begin{equation}
\widehat{x}_{i}=\left\langle \pi\left(  x\right)  \left\vert \widehat{x}%
_{i}\right\vert \phi\left(  y\right)  \right\rangle _{n}=\frac{\hbar}{2}%
x_{i}{\delta_{n}\left(  x-y\right)  }%
\end{equation}
where $x_{i}$ is on the $n$ hyperplane.

The relationship of the relativistic position operator to the nonrelativistic
position operator $\mathrm{i}\mathbf{\nabla}_{\mathbf{k}}$ and the NW position
operator can be seen by transforming to Fourier space. The position operator
(\ref{PositionOperator}) is in the IP while the conventional position operator
is time independent so it is in the Schr\"{o}dinger picture (SP). The unitary
time evolution operator that transforms states and operators from the SP to
the IP is, from (\ref{t_dependence}), $\widehat{U}\left(  t\right)
=\exp\left(  -{\mathrm{i}}\epsilon\widehat{D}^{1/2}t\right)  .$ Using
$\widehat{\mathbf{x}}^{{\mathrm{SP}}}=\widehat{U}\widehat{\mathbf{x}%
}\widehat{U}^{\dagger}$and expanding in the biorothogonal bases as in
(\ref{F}), the positive frequency part of the SP position operator
(\ref{PositionOperator}) is
\begin{multline}
\widehat{\mathbf{x}}^{{\mathrm{SP}}+}=\int\frac{\mathrm{d}\mathbf{k}}{2\left(
2\pi\right)  ^{3}}\int\frac{\mathrm{d}\mathbf{q}}{2\left(  2\pi\right)  ^{3}%
}\\
\int{\mathrm{d}\mathbf{x\ x}\,\mathrm{e}}^{-\mathrm{i}\mathbf{k}%
\cdot\mathbf{x}}\left\vert \frac{1_{\mathbf{k}}}{\omega_{\mathbf{k}}%
}\right\rangle \left\langle 1_{\mathbf{q}}\right\vert {\mathrm{e}}%
^{\mathrm{i}\mathbf{q}\cdot\mathbf{x}}. \label{k_pos_op}%
\end{multline}
Since $\int{\mathrm{d}\mathbf{x}}\,{\mathrm{e}}^{\mathrm{i}\mathbf{q}%
\cdot\mathbf{x}}\mathrm{i}\mathbf{\nabla}_{\mathbf{k}}{\mathrm{e}%
}^{-\mathrm{i}\mathbf{k}\cdot\mathbf{x}}=2\left(  2\pi\right)  ^{3}%
\mathrm{i}\mathbf{\nabla}_{\mathbf{k}}\delta\left(  \mathbf{q}-\mathbf{k}%
\right)  $ \cite{Sakurai},%
\begin{align}
\widehat{\mathbf{x}}^{{\mathrm{SP}}+}  &  =\int\frac{\mathrm{d}\mathbf{k}%
}{2\left(  2\pi\right)  ^{3}}\left\vert \frac{1_{\mathbf{k}}}{\omega
_{\mathbf{k}}}\right\rangle \mathrm{i}\mathbf{\nabla}_{\mathbf{k}}\left\langle
1_{\mathbf{k}}\right\vert ,\\
\widehat{\mathbf{x}}^{{\mathrm{SP}}+}\left(  \mathbf{k}\right)   &
=\mathrm{i}\mathbf{\nabla}_{\mathbf{k}}, \label{pos_op}%
\end{align}
so that $\mathrm{i}\mathbf{\nabla}_{\mathbf{k}}$ is the position operator in
the $\left\vert 1_{\mathbf{k}}/\omega_{\mathbf{k}}\right\rangle \left\langle
1_{\mathbf{k}}\right\vert $ basis. When operating on $\mathrm{e}%
^{-\mathrm{i}\mathbf{k}\cdot\mathbf{x}}$ it extracts the position $\mathbf{x}$
where the particle was created. With positive definite Hermian metric $\Theta$
the scalar product is $\left\langle .|\Theta.\right\rangle $ and an operator
$\widehat{O}$ is quasi-Hermitian if $\Theta\widehat{O}^{\dagger}%
=\widehat{O}\Theta$. Since according to (\ref{PositionOperator})
$\widehat{\mathbf{x}}^{\dagger}=\widehat{D}^{1/2}\widehat{\mathbf{x}%
}\widehat{D}^{-1/2}$, the position operator is quasi-Hermitian
\cite{GeyerScholtz}. Defining $S\equiv\Theta^{1/2},$ the operator
$\widehat{o}=\widehat{o}^{\dagger}=S^{-1}\widehat{O}^{\dagger}S$ is Hermitian
\cite{GeyerScholtz}. In (\ref{k_sp}) the metric is $\Theta=\omega_{\mathbf{k}%
}^{-1}$ so $S=\omega_{\mathbf{k}}^{-1/2},$ the basis is $\left\{
\omega_{\mathbf{k}}^{-1/2}\left\vert 1_{\mathbf{k}}\right\rangle \right\}  $
and the matrix representing the NW position operator
\cite{Mostafazadeh1,NewtonWigner,Mostafazadeh2},%
\begin{equation}
\widehat{\mathbf{x}}_{{\mathrm{NW}}}^{{\mathrm{SP}}}\left(  \mathbf{k}\right)
=\omega_{\mathbf{k}}^{1/2}\mathrm{i}\mathbf{\nabla}_{\mathbf{k}}%
\omega_{\mathbf{k}}^{-1/2},
\end{equation}
is of the form $\widehat{o}^{\dagger}=S^{-1}\widehat{O}^{\dagger}S$ for
$\widehat{O}^{\dagger}=\widehat{\mathbf{x}}^{{\mathrm{SP}}+}\left(
\mathbf{k}\right)  $ where $\omega_{\mathbf{k}}^{1/2}\mathrm{i}\mathbf{\nabla
}_{\mathbf{k}}\omega_{\mathbf{k}}^{-1/2}=\mathrm{i}\mathbf{\nabla}%
_{\mathbf{k}}-\frac{\mathrm{i}c^{2}}{\omega_{\mathbf{k}}^{2}}\mathbf{k}$. The
factors $\omega_{\mathbf{k}}^{\pm1/2}$ introduce nonlocality into the
configuration space description of the position eigenvectors. These nonlocal
factors also appear in the expressions for the field operators in quantum
optics \cite{ScullyZubairy,CohenQED1,HawtonBaylis}, but we see here that this
nonlocality is not physical since it does not appear in the covariant
description of the position observable.

\section{Photons}

\label{sec:Photons}

For photons the scalar field $\phi$ should be replaced with the four-vector
potential $A^{\mu}$. Both $A_{\mu}\partial^{\nu}A^{\mu}=\left(  A_{\mu
}\partial_{ct}A^{\mu},A_{\mu}\nabla A^{\mu}\right)  $ and $A_{\nu}F^{\nu\mu
}=\left(  \mathbf{A}\cdot\mathbf{E}/c,\mathbf{A}\times\mathbf{B}\right)  $
multiplied by $\mathrm{i}\epsilon_{0}c/\hbar$ are candidates for the
four-current density, $F^{\nu\mu}=\partial^{\nu}A^{\mu}-\partial^{\mu}A^{\nu}$
being the Faraday tensor. The properties of an operator of the form
$\mathrm{i}A_{\nu}F^{\nu\mu}$ were investigated in \cite{NumDens95}. The
Coulomb gauge condition $\nabla\cdot\mathbf{A}=0$ is not Lorentz invariant,
but $A_{\mu}$ can be chosen to transform as a Lorentz four-vector up to an
extra term that maintains the Coulomb gauge in all frames of reference
\cite{Debierre}. To avoid the complications associated with nonphysical
longitudinal and scalar photons the Coulomb gauge will be used here. In a
source-free region in the Coulomb gauge both $A_{\mu}\partial^{\nu}A^{\mu}$
and $A_{\nu}F^{\nu\mu}$ reduce to $\left(  \mathbf{A}_{\perp}\cdot
\mathbf{E}_{\perp}/c,\mathbf{A}_{\perp}\times\mathbf{B}\right)  $.

Following (\ref{M}) we can define%
\begin{equation}
J^{\mu}\left(  x\right)  =\frac{\mathrm{i}\epsilon_{0}c}{\hbar}\sum
_{\lambda,i}A_{\lambda i}\left(  x\right)  ^{\ast}\overleftrightarrow{\partial
}^{\mu}A_{\lambda ci}\left(  x\right)  \label{Jph}%
\end{equation}
where\textbf{ }$\mathbf{A}_{\lambda}$\textbf{ }is a transverse vector
potential of helicity\textbf{ }$\lambda$ that satisfies the classical Maxwell
wave equation {$\left(  \widehat{D}+\partial_{ct}^{2}\right)  \mathbf{A}%
_{\lambda}=0$ }where\textbf{ }$m=0$\textbf{ }so{ $\widehat{D}=-\nabla^{2}.$
The} conjugate field is\textbf{ }$\mathbf{A}_{\lambda c}\equiv\mathrm{i}%
\widehat{D}^{-1/2}\partial_{ct}\mathbf{A}_{\lambda}=\mathbf{A}_{\lambda}%
^{+}-\mathbf{A}_{\lambda}^{-}$ where\textbf{ }$\mathbf{A}_{\lambda}%
=\mathbf{A}_{\lambda}^{+}+\mathbf{A}_{\lambda}^{-}.$\textbf{ }{It can then be
verified by substitution that (\ref{Jph}) satisfies the continuity equation
$\partial_{\mu}J^{\mu}\left(  x\right)  =0$. As a consequence the scalar
product $\int_{\sigma}\mathrm{d}\sigma n_{\mu}J^{\mu}\left(  x\right)  $ is
Lorentz invariant. T{he photon scalar product that replaces (\ref{sp}) is%
\begin{equation}
\left(  A_{1},A_{2}\right)  =\frac{2\epsilon_{0}c}{\hbar}\sum_{\lambda
,\epsilon,i}\left\langle A_{1\lambda i}^{\epsilon}|\widehat{D}^{1/2}%
A_{2\lambda i}^{\epsilon}\right\rangle \label{photon_sp}%
\end{equation}
where }$A=A^{\mu}=\left(  0,\mathbf{A}_{\perp}\right)  $. }

For photons described in the Coulomb gauge, the field operator is
$\widehat{\mathbf{A}}_{\perp}\left(  \mathbf{x},t\right)  $ and its canonical
conjugate is $-\epsilon_{0}\widehat{\mathbf{E}}_{\perp}\left(  \mathbf{x}%
,t\right)  $ where $\widehat{\mathbf{E}}_{\perp}=-\partial_{t}%
\widehat{\mathbf{A}}_{\perp}$ is the electric field operator. The Fourier
space spherical polar coordinates will be called $k,$ $\theta_{\mathbf{k}}$
and $\phi_{\mathbf{k}}$ and their corresponding unit vectors $\mathbf{e}%
_{\mathbf{k}},$ $\mathbf{e}_{\theta}$ and $\mathbf{e}_{\phi}.$ The definite
helicity transverse unit vectors are $\mathbf{e}_{\lambda}\left(
\mathbf{k}\right)  =\left(  \mathbf{e}_{\theta}+\mathrm{i}\lambda
\mathbf{e}_{\phi}\right)  /\sqrt{2}$ where $\lambda$ is helicity, and the
longitudinal unit vector is $\mathbf{e}_{\mathbf{k}}$.{ }The NW photon
position operator with commuting components can be written in Fourier space as
\cite{HawtonBaylis} $\widehat{\mathbf{x}}=\widehat{E}\left(  \mathrm{i}%
\omega_{\mathbf{k}}^{1/2}\nabla_{\mathbf{k}}\omega_{\mathbf{k}}^{-1/2}\right)
\widehat{E}^{-1}$ where $\widehat{E}$ is a rotation through Euler angles to
fixed reference axes. The basic idea is the same as was used in the derivation
of the NW position operator in \cite{Mostafazadeh1}; the position information
is contained in the factor $\exp\left(  \mathrm{i}\mathbf{k\cdot x}\right)  $
in the wave function but the factors $\omega_{\mathbf{k}}^{\pm1/2}$ must be
eliminated before the nonrelativistic position operator $\mathrm{i}%
\nabla_{\mathbf{k}}$ can be used to extract this information. For the
transverse fields that describe photons an additional unitary transformation
$\widehat{E}$ that rotates the field vectors to axes fixed in space is needed.
A position eigenvector has a vortex structure like twisted light
\cite{Twisted,HawtonBaylis} with nonzero spatial extension in which the photon
position eigenvalue $\mathbf{x}$ is the center of internal angular momentum
\cite{Fleming}. Here the NW-like position operator derived in
\cite{HawtonPosOp} will not be used, but the definite helicity basis vectors
$e_{\lambda}\left(  \mathbf{k}\right)  $ that are defined for all $\mathbf{k}$
are still needed to describe position eigenvectors.

Following the derivation in Section \ref{sec:Scalar}, the IP photon position
operator is%
\begin{equation}
\widehat{\mathbf{x}}=\frac{2}{\hbar}\sum_{\epsilon,\lambda=\pm}\int%
\mathrm{d}\mathbf{x\ x}\left\vert \mathbf{A}_{\lambda}^{\epsilon}\left(
x\right)  \right\rangle \cdot\left\langle \mathbf{E}_{\lambda}^{\epsilon
}\left(  x\right)  \right\vert , \label{x_photon}%
\end{equation}
the position eigenvector equations are
\begin{subequations}
\label{photon_evecs}%
\begin{align}
\widehat{\mathbf{x}}\left\vert \mathbf{A}_{\lambda}^{\epsilon}\left(
x\right)  \right\rangle  &  =\mathbf{x}\left\vert \mathbf{A}_{\lambda
}^{\epsilon}\left(  x\right)  \right\rangle ,\\
\widehat{\mathbf{x}}^{\dagger}\left\vert \mathbf{E}_{\lambda}^{\epsilon
}\left(  x\right)  \right\rangle  &  =\mathbf{x}\left\vert \mathbf{E}%
_{\lambda}^{\epsilon}\left(  x\right)  \right\rangle
\label{photon_eivenvectors}%
\end{align}
and position basis states are
\end{subequations}
\begin{subequations}
\label{AandE}%
\begin{align}
\left\vert \mathbf{A}_{\lambda}\left(  x\right)  \right\rangle  &
\equiv\widehat{\mathbf{A}}_{\lambda}^{-}\left(  x\right)  \left\vert
0\right\rangle ,\label{A_loc}\\
\left\vert \mathbf{E}_{\lambda}\left(  x\right)  \right\rangle  &  \equiv
c\widehat{D}^{1/2}\widehat{\mathbf{A}}_{\lambda}^{-}\left(  x\right)
\left\vert 0\right\rangle . \label{E_loc}%
\end{align}
Here the potential operator reads
\end{subequations}
\begin{equation}
\widehat{\mathbf{A}}_{\lambda}^{-}\left(  x\right)  =\sqrt{\frac{\hbar
}{\epsilon_{0}}}\int\frac{\mathrm{d}\mathbf{k}}{\left(  2\pi\right)
^{3}2\omega_{\mathbf{k}}}\mathbf{e}_{\lambda}\left(  \mathbf{k}\right)
\mathrm{e}^{-\mathrm{i}\left(  \mathbf{k}\cdot\mathbf{x}-\omega_{\mathbf{k}%
}t\right)  }\hat{a}_{\lambda}^{\dagger}\left(  \mathbf{k}\right)  . \label{A}%
\end{equation}
The Fourier space canonical commutation relations and orthogonality relations
are
\begin{subequations}
\label{k_normalization}%
\begin{align}
\left[  \widehat{a}_{\lambda}\left(  \mathbf{k}\right)  ,\widehat{a}_{\sigma
}^{\dagger}\left(  \mathbf{q}\right)  \right]   &  =\left(  2\pi\right)
^{3}2\omega_{\mathbf{k}}\,\delta\left(  \mathbf{k}-\mathbf{q}\right)
\delta_{\lambda\sigma},\label{photon_k_commutation}\\
\left\langle 1_{\lambda,\mathbf{k}}|1_{\sigma,\mathbf{q}}\right\rangle  &
=\left(  2\pi\right)  ^{3}2\omega_{\mathbf{k}}\,\delta\left(  \mathbf{k}%
-\mathbf{q}\right)  \delta_{\lambda\sigma}, \label{photon_k_normalization}%
\end{align}
where $\left\vert 1_{\lambda,\mathbf{k}}\right\rangle \equiv a_{\lambda
}^{\dagger}\left(  \mathbf{k}\right)  \left\vert 0\right\rangle .$

With $\epsilon=\pm$ states defined as in Section \ref{sec:Scalar}, the photon
position eigenvectors are biorthogonal since their commutation relations imply
that
\end{subequations}
\begin{equation}
\sum_{i=1}^{3}\left\langle E_{\lambda i}^{\epsilon}\left(  x\right)
|A_{\sigma i}^{\epsilon^{\prime}}\left(  y\right)  \right\rangle =\frac{\hbar
}{2\epsilon_{0}}{\delta_{n}\left(  x-y\right)  }\delta_{\lambda\sigma}%
\delta_{\epsilon\epsilon^{\prime}} \label{orthogonal}%
\end{equation}
where the subscripts $i$ denote Cartesian components of the three-vectors
and\textbf{ }$\epsilon=+$ for absorption at $x$ while $\epsilon=-$ for
emission at $x$. For these position eigenvectors the scalar product
(\ref{photon_sp}) is\textbf{ }$\left(  A_{\lambda}^{\epsilon}\left(  x\right)
,A_{\sigma}^{\epsilon}\left(  y\right)  \right)  ={\delta_{n}\left(
x-y\right)  }\delta_{\lambda\sigma}.$ For free photons described by transverse
fields the completeness relation is%
\begin{equation}
\widehat{1}_{\perp}=\frac{2\epsilon_{0}}{\hbar}\sum_{\epsilon,\lambda=\pm}%
\int\mathrm{d}\mathbf{x}\left\vert \mathbf{A}_{\lambda}^{\epsilon}\left(
x\right)  \right\rangle \cdot\left\langle \mathbf{E}_{\lambda}^{\epsilon
}\left(  x\right)  \right\vert \label{complete}%
\end{equation}
where we have defined%
\begin{equation}
\left\vert \mathbf{A}_{\lambda}^{\epsilon}\left(  x\right)  \right\rangle
\cdot\left\langle \mathbf{E}_{\lambda}^{\epsilon}\left(  x\right)  \right\vert
\equiv\sum_{i=1}^{3}\left\vert A_{\lambda i}^{\epsilon}\left(  x\right)
\right\rangle \left\langle E_{\lambda i}^{\epsilon}\left(  x\right)
\right\vert . \label{dot}%
\end{equation}
The identity operator $\widehat{1}_{\perp}$ on the space of transverse photons
is closely connected to the so-called `transverse Dirac delta' of QED
\cite{CohenQED1}. For any transverse one-photon state we can thence write
\begin{equation}
\left\vert \psi_{\perp}\left(  t\right)  \right\rangle =\frac{2\epsilon_{0}%
}{\hbar}\sum_{\epsilon,\lambda=\pm}\int\mathrm{d}\mathbf{x}\left\vert
\mathbf{A}_{\lambda}^{\epsilon}\left(  x\right)  \right\rangle \cdot
\left\langle \mathbf{E}_{\lambda}^{\epsilon}\left(  x\right)  |\psi^{\epsilon
}\left(  t\right)  \right\rangle \label{one_photon}%
\end{equation}
and the wave function
\begin{equation}
\mathbf{\psi}_{\lambda}^{\epsilon}\left(  x\right)  =\left\langle
\mathbf{E}_{\lambda}^{\epsilon}\left(  x\right)  |\psi^{\epsilon}\left(
t\right)  \right\rangle =\left\langle \mathbf{A}_{\lambda}^{\epsilon}\left(
x\right)  |\widetilde{\psi}^{\epsilon}\left(  t\right)  \right\rangle
\label{wave_function}%
\end{equation}
completely describes the one-photon state $\left\vert \psi_{\perp}\left(
t\right)  \right\rangle $ in either basis of position eigenvectors. The dual
state vector is
\begin{equation}
\left\vert \widetilde{\psi}_{\perp}\left(  t\right)  \right\rangle
=\frac{2\epsilon_{0}}{\hbar}\sum_{\epsilon,\lambda=\pm}\int\mathrm{d}%
\mathbf{x}\left\vert \mathbf{E}_{\lambda}^{\epsilon}\left(  x\right)
\right\rangle \cdot\left\langle \mathbf{A}_{\lambda}^{\epsilon}\left(
x\right)  |\widetilde{\psi}^{\epsilon}\left(  t\right)  \right\rangle ,
\end{equation}
the squared norm is
\begin{equation}
\left\langle \psi_{\perp}|\widetilde{\psi}_{\perp}\right\rangle =\frac
{2\epsilon_{0}}{\hbar}\sum_{\epsilon,\lambda=\pm}\int\mathrm{d}\mathbf{x}%
\left\vert \mathbf{\psi}_{\lambda}^{\epsilon}\left(  x\right)  \right\vert
^{2}%
\end{equation}
and the \emph{probability density }per photon for a transition from
$\left\vert \psi_{\perp}\left(  t\right)  \right\rangle $ to the $\epsilon
$-frequency position eigenvector with helicity $\lambda$%
\begin{align}
p_{\lambda}^{\epsilon}\left(  x\right)   &  =\frac{2\epsilon_{0}}%
{\hbar\left\langle \psi|\widetilde{\psi}\right\rangle }\left\vert
\mathbf{\psi}_{\lambda}^{\epsilon}\left(  x\right)  \right\vert ^{2}%
\label{probability}\\
&  =\frac{\left\vert \mathbf{\psi}_{\lambda}^{\epsilon}\left(  x\right)
\right\vert ^{2}}{\sum_{\epsilon,\lambda=\pm}\int\mathrm{d}\mathbf{x}%
\left\vert \mathbf{\psi}_{\lambda}^{\epsilon}\left(  x\right)  \right\vert
^{2}}\nonumber
\end{align}
is positive definite.

A one-photon state can be Fourier expanded as%
\begin{equation}
\left\vert \psi_{\perp}\left(  t\right)  \right\rangle =\sum_{\lambda
,\epsilon=\pm}\int\frac{\mathrm{d}\mathbf{k}}{\left(  2\pi\right)  ^{3}%
2\omega_{\mathbf{k}}}c_{\lambda}^{\epsilon}\left(  \mathbf{k},t\right)
\left\vert 1_{\lambda,\mathbf{k}}\right\rangle . \label{k_state}%
\end{equation}
Eqs. (\ref{AandE}), (\ref{A}) and (\ref{photon_k_normalization}) give\textbf{
}$\left\langle \mathbf{E}_{\lambda}\left(  \mathbf{x}\right)  |1_{\lambda
,\mathbf{k}}\right\rangle /\omega_{\mathbf{k}}=\left\langle \mathbf{A}%
_{\lambda}\left(  \mathbf{x}\right)  |1_{\lambda,\mathbf{k}}\right\rangle $ so
that $\left\vert 1_{\lambda,\mathbf{k}}/\omega_{\mathbf{k}}\right\rangle
\in\mathcal{H}$\ and\textbf{ }$\left\vert 1_{\lambda,\mathbf{k}}\right\rangle
\in\mathcal{H}^{\ast}.$\textbf{ }Substitution in (\ref{wave_function}) then
gives the dual state vector%
\begin{equation}
\left\vert \widetilde{\psi}_{\perp}\left(  t\right)  \right\rangle
=\sum_{\lambda,\epsilon=\pm}\int\frac{\mathrm{d}\mathbf{k}}{\left(
2\pi\right)  ^{3}2}c_{\lambda}^{\epsilon}\left(  \mathbf{k},t\right)
\left\vert 1_{\lambda,\mathbf{k}}\right\rangle . \label{k_dual}%
\end{equation}
The probability amplitude for a transition to a $\epsilon$-frequency plane
wave state with wave vector\textbf{ }$\mathbf{k}$ and helicity\textbf{
}$\lambda$\textbf{ }is proportional to\textbf{ }$\left\langle 1_{\lambda
,\mathbf{k}}|\psi^{\epsilon}\left(  t\right)  \right\rangle =\left\langle
1_{\lambda,\mathbf{k}}/\omega_{\mathbf{k}}|\widetilde{\psi}^{\epsilon}\left(
t\right)  \right\rangle =c_{\lambda}^{\epsilon}\left(  \mathbf{k},t\right)  .$
According to the rules of biorthogonal QM outlined in the Introduction the
probability density per photon for this transition is%
\begin{align}
p_{\lambda}^{\epsilon}\left(  \mathbf{k}\right)   &  =\frac{\left\vert
\left\langle 1_{\lambda,\mathbf{k}}|\psi^{\epsilon}\left(  t\right)
\right\rangle \right\vert ^{2}}{\left\langle \widetilde{\psi}\left(  t\right)
|\psi\left(  t\right)  \right\rangle \left(  2\pi\right)  ^{3}2}\nonumber\\
&  =\frac{\left\vert c_{\lambda}^{\epsilon}\left(  \mathbf{k},t\right)
\right\vert ^{2}}{\sum_{\lambda,\epsilon=\pm}\int\mathrm{d}\mathbf{k}%
\,\left\vert c_{\lambda}^{\epsilon}\left(  \mathbf{k},t\right)  \right\vert
^{2}}. \label{p_k}%
\end{align}
Time dependence of\textbf{ }$c_{\lambda}^{\epsilon}\left(  \mathbf{k}%
,t\right)  $ indicates the presence of a source. When a photon is emitted by
an atom, the expectation value of the photon number is smaller than one and
approaches unity as $t\rightarrow\infty.$\textbf{ }If\textbf{ }$\left\vert
\psi_{\perp}\left(  t\right)  \right\rangle $ \textbf{i}s normalized so
that\textbf{ }$n\left(  t\right)  =\left\langle \widetilde{\psi}\left(
t\right)  |\psi\left(  t\right)  \right\rangle $ is the number of photons, the
probability density for\textbf{ }$k^{\mu}=\left(  \epsilon\omega_{\mathbf{k}%
},\mathbf{k}\right)  $ is\textbf{ }$\left\vert c_{\lambda}^{\epsilon}\left(
\mathbf{k},t\right)  \right\vert ^{2}/\left[  \left(  2\pi\right)
^{3}2\right]  $\textbf{ }while the probability density to find a\textbf{
}photon\textbf{ }at \textbf{$x$ }on the hyperplane $\sigma$ is $\left\vert
\mathbf{\psi}_{\lambda}^{\epsilon}\left(  x\right)  \right\vert ^{2}%
2\epsilon_{0}/\hbar$. In the SP the photon position operator (\ref{x_photon})
can be written as%
\begin{equation}
\widehat{\mathbf{x}}^{{\mathrm{SP}}}\left(  \mathbf{k}\right)  =\widehat{E}%
\mathrm{i}\mathbf{\nabla}_{\mathbf{k}}\widehat{E}^{-1}.
\end{equation}

The scalar product\textbf{ }$\left\langle \mathbf{E}_{\lambda}^{+}\left(
x\right)  |\psi^{+}\left(  t\right)  \right\rangle =\left\langle
\mathbf{E}_{\lambda}\left(  x\right)  |\psi\left(  t\right)  \right\rangle $
that leads to an invariant probability to count a photon is proportional to
probability amplitude, not the electric field. {Glauber defined an ideal
photon detector as a system of negligible size with a frequency-independent
photon absorption probability \cite{Glauber}. For the positive frequency
one-photon state $\left\vert \psi\left(  t\right)  \right\rangle $ he found
that the probability to count a photon is proportional to }$\left\vert
\left\langle \mathbf{E}_{\lambda}\left(  x\right)  |\psi\left(  t\right)
\right\rangle \right\vert ^{2}${$.{\ }$Glauber considered photodetection to be
a square law process and interpreted it to be responsive to the density of
electromagnetic energy, but }number density gives an invariant probability to
count a photon while energy density does not.{ Indeed, the biorthogonal
completeness relation (\ref{complete}) implies that a basis of ideal Glauber
detectors can be defined provided the state vector $\left\vert \psi
\right\rangle \in\mathcal{H}$ of the photon at hand has been created by the
$\mathbf{A}\cdot\mathbf{p}$ minimal coupling Hamiltonian. In that case,
$\left\vert \left\langle \mathbf{E}_{\lambda}\left(  x\right)  |\psi\left(
t\right)  \right\rangle \right\vert ^{2}$ is proportional to photon
probability density.}

{ } Here a positive definite particle density is obtained in the physical
Hilbert space according to the rules of biorthogonal QM summarized in the
Introduction. An alternative approach is to transform the physical fields to{
the Foldy representation \cite{Foldy} using the nonlocal operator
}$\widehat{D}^{-1/4}$ and its inverse{ {\cite{Mostafazadeh1}}. For photons
this nonlocal transformation leads to the Landau-Peierls (LP) wave function
\cite{BB96,LandauPeierls}. The disadvantage to this approach is that the
relationship between the LP wave function and a current source is nonlocal.
Here the fields }$A^{\mu}\left(  x\right)  ${ due to the local interaction
Hamilton }$j_{\mu}\left(  x\right)  A^{\mu}\left(  x\right)  ${ are calculated
first and the probability amplitude for a transition to a position eigenvector
is then obtained using the invariant scalar product (\ref{photon_sp}). These
fields have well defined Lorentz and gauge transformation properties. The
positive definiteness of the probability follows then directly from the
mathematical rules of biorthogonal QM.}

An advantage of the second quantized formalism used here is that multiphoton
wave functions can be introduced as in \cite{Glauber,SmithRaymer,Hawton07}.
For example, to the two photon state $\left\vert \psi_{2}\right\rangle $ can
be associated the wave function%
\begin{equation}
\psi_{\lambda_{1},\lambda_{2}}\left(  \mathbf{x}_{1},\mathbf{x}_{2},t\right)
=\left\langle \mathbf{E}_{\lambda_{1}}\left(  \mathbf{x}_{1}\right)
\mathbf{E}_{\lambda_{2}}\left(  \mathbf{x}_{2}\right)  |\psi_{2}\left(
t\right)  \right\rangle
\end{equation}
with $\left\vert \mathbf{E}_{\lambda_{1}}\left(  \mathbf{x}_{1}\right)
\mathbf{E}_{\lambda_{2}}\left(  \mathbf{x}_{2}\right)  \right\rangle
\equiv\widehat{\mathbf{E}}_{\lambda_{1}}\left(  \mathbf{x}_{1}\right)
\widehat{\mathbf{E}}_{\lambda_{2}}\left(  \mathbf{x}_{2}\right)  \left\vert
0\right\rangle .$ This wave function localizes the photons at spatial points
$\mathbf{x}_{1}$ and $\mathbf{x}_{2}$ at time $t$ and can describe entangled
two-photon states.

\section{Wave function of a photon emitted by an atom}

\label{sec:Emission}

The wave function (\ref{wave_function}) for a photon emitted by a two-level
atom initially in its excited state was derived in \cite{Sipe} and
\cite{Debierre,DebierreDurt}, to{ first order in the IP minimal coupling
interaction Hamiltonian $\widehat{H}_{I}=\left(  e/m_{e}\right)
\widehat{\mathbf{A}}\left(  \hat{\mathbf{x}}_{e},t\right)  \cdot
\hat{\mathbf{p}}_{e}$}$\left(  t\right)  .$ For a two-level atom initially in
its excited state $\left\vert \mathrm{e}\right\rangle $ with no photons
present, the positive frequency IP wave function describing decay to its
ground state $\left\vert \mathrm{g}\right\rangle $ while emitting one photon
is%
\begin{align}
\left\vert \psi\left(  t\right)  \right\rangle  &  =c_{\mathrm{e}}\left(
t\right)  \left\vert \mathrm{e},0\right\rangle \label{eq:IPState}\\
&  +\sum_{\lambda=\pm}\int\frac{\mathrm{d}\mathbf{k}}{\left(  2\pi\right)
^{3}2\omega_{\mathbf{k}}}c_{\mathrm{g},\lambda}\left(  \mathbf{k},t\right)
\left\vert \mathrm{g},1_{\lambda,\mathbf{k}}\right\rangle \nonumber
\end{align}
where
\begin{align}
c_{\mathrm{g},\lambda}\left(  \mathbf{k},t\right)   &  =\frac{e}{m_{e}%
}\left\langle \mathrm{g},1_{\lambda,\mathbf{k}}\left\vert \widehat{\mathbf{A}%
}_{\lambda}^{-}\left(  \hat{\mathbf{x}}_{e},t\right)  \cdot\hat{\mathbf{p}%
}_{e}\left(  t\right)  \right\vert \mathrm{e},0\right\rangle
\label{eq:GroundPhoton}\\
&  \times\frac{1-\mathrm{e}^{\mathrm{i}\left(  \omega_{\mathbf{k}}-\omega
_{0}\right)  t}}{\hbar\left(  \omega_{\mathbf{k}}-\omega_{0}\right)
}.\nonumber
\end{align}
Here $\hbar\omega_{0}$ is the level separation between the ground and excited
states and $\hat{\mathbf{x}}_{e}$ and $\hat{\mathbf{p}}_{e}$ are the electron
position and momentum operators. For $\left\vert \widetilde{\psi}_{\perp
}\right\rangle $ given by (\ref{k_dual}) the transverse single-photon state
and its dual are thus given by \label{eq:PsiPsiTilde}
\begin{subequations}
\label{phi&psitilde}%
\begin{align}
\left\vert \psi_{\perp}\left(  t\right)  \right\rangle  &  =\sum_{\lambda=\pm
}\int\frac{\mathrm{d}\mathbf{k}}{\left(  2\pi\right)  ^{3}2\omega_{\mathbf{k}%
}}c_{\mathrm{g},\lambda}\left(  \mathbf{k},t\right)  \left\vert 1_{\lambda
,\mathbf{k}}\right\rangle ,\\
\left\vert \widetilde{\psi}_{\perp}\left(  t\right)  \right\rangle  &
=\sum_{\lambda=\pm}\int\frac{\mathrm{d}\mathbf{k}}{\left(  2\pi\right)  ^{3}%
2}c_{\mathrm{g},\lambda}\left(  \mathbf{k},t\right)  \left\vert 1_{\lambda
,\mathbf{k}}\right\rangle .
\end{align}

In \cite{Debierre,DebierreDurt} the minimal coupling Hamiltonian created the
photon state vector $\left\vert \psi_{\perp}\right\rangle $ in the $\left\vert
\mathbf{A}_{\lambda}\left(  \mathbf{x}\right)  \right\rangle $ basis so the
appropriate wave function is $\left\langle \mathbf{E}_{\lambda}\left(
\mathbf{x}\right)  |\psi_{\perp}\right\rangle $. Indeed we have from
(\ref{wave_function}) and {(\ref{eq:IPState})}%
\end{subequations}
\begin{equation}
\mathbf{\psi}_{\lambda}\left(  x\right)  =\mathrm{i}\sqrt{\frac{\hbar
}{\epsilon_{0}}}\int\frac{\mathrm{d}\mathbf{k}}{\left(  2\pi\right)  ^{3}%
2}\mathbf{e}_{\lambda}\left(  \mathbf{k}\right)  \mathrm{e}^{-\mathrm{i}%
k_{\mu}x^{\mu}}c_{\mathrm{g},\lambda}\left(  \mathbf{k},t\right)  .
\label{atom_wf}%
\end{equation}
The positive frequency wave function of the emitted photon is calculated, but
causal solutions that include negative frequencies are also considered. Taking
into account a factor $-\mathrm{i}$ in the electric field operator used in
\cite{Debierre,DebierreDurt}, $c_{\mathrm{g},\lambda}=-\mathrm{i}c_{\lambda
}^{+}$ in (\ref{k_state})\textbf{.} Substitution of the wave function
$\mathbf{\psi}_{\lambda}\left(  x\right)  =\left\langle \mathbf{E}_{\lambda
}\left(  \mathbf{x}\right)  |\psi_{\perp}\left(  t\right)  \right\rangle $ in
(\ref{probability}) gives the probability density in space to count a photon
at time $t.$\textbf{ }Since in \cite{Debierre,DebierreDurt} the wave function
is normalized as $\left\langle \psi_{\perp}|\psi_{\perp}\right\rangle =1,$ the
factor $\left\langle \psi_{\perp}|\widetilde{\psi}_{\perp}\right\rangle $ in
(\ref{probability}) and (\ref{p_k}) approaches\textbf{ }$1/\omega_{0}$ as $t$
$\rightarrow\infty.$

If the standard (dipolar) $\mathbf{E}\cdot\mathbf{x}$ Hamiltonian were to be
used instead, the photon would be created in the $\left\vert \mathbf{E}%
_{\lambda}\left(  x\right)  \right\rangle $\ basis and the appropriate wave
function would be $\left\langle \mathbf{A}_{\lambda}\left(  x\right)
|\widetilde{\psi}_{\perp}\left(  t\right)  \right\rangle .$

\section{Conclusion}

\label{sec:Ccl}

The formalism of biorthogonal systems can be, as we saw, called in action in
relativistic quantum mechanics. It is particularly well-matched to the
relativistic scalar product. In the biorthogonal formalism, both the
Wigner-Bargmann quantum field operator (for photons, the vector potential) and
its canonically conjugate momentum (for photons, the electric field) are put
on an equal footing, and they generate respectively the direct and the dual
basis of position eigenvectors of two different position operators, which are
the Hermitian conjugate of each other. Our formalism further clarifies the
meaning of the free parameter $\alpha$ \cite{HawtonBaylis,Hawton07,Debierre}
in the photon position operator. Here, the freedom in the choice of $\alpha$
allows us to use both $\mathbf{x}$ and $\mathbf{x}^{\dagger}$. \newline

The probability density (\ref{probability}) suggests a resolution of the
apparent dichotomy between photon number counting and the sensitivity of a
detector to energy density. The wave function $\left\langle \mathbf{E}%
_{\lambda}^{+}\left(  x\right)  |\psi\left(  t\right)  \right\rangle $
together with the state vector (\ref{eq:IPState}) describes creation of a
photon in the time interval $0\leq t^{\prime}\leq$ $t$ followed by its
detection at time $t.$ Since it is created in the $\left\vert \mathbf{A}%
_{\lambda}^{\epsilon}\left(  x\right)  \right\rangle $ basis and observed in
the dual $\left\vert \mathbf{E}_{\lambda}^{\epsilon}\left(  x\right)
\right\rangle $ basis, that wave function is proportional to a probability
amplitude. The probability density for a transition from $\left\vert
\psi_{\perp}\left(  t\right)  \right\rangle $ to the position eigenvector at
$\mathbf{x}$, given by $2\epsilon_{0}\left\vert \left\langle \mathbf{E}%
_{\lambda}^{+}\left(  x\right)  |\psi\left(  t\right)  \right\rangle
\right\vert ^{2}/\hbar,$ is of the Glauber form \cite{Glauber}. However, in
contrast to theories of photodetection based on energy density, we have
proposed, through the position amplitude $\left\langle \mathbf{E}_{\lambda
}^{+}\left(  x\right)  |\psi\left(  t\right)  \right\rangle $ in the dual
basis, a true position measurement that describes an array of ideal photon
counting detectors.

{For a state vector that is an arbitrary linear combination of positive and
negative frequency terms the probability density for a transition to the
position eigenvector at $\mathbf{x}$ is positive definite. This particle plus
antiparticle probability density describes a particle at spatial location
$\mathbf{x}$ independent of whether it was absorbed or emitted. Thus
(\ref{probability}) can be interpreted as probability density even if the wave
function (\ref{wave_function}) is real as in classical electromagnetism. }This
application of biorthogonal QM is based on an invariant positive definite
scalar product so transition probabilities are invariant and positive
definite, the position operator is covariant, and there is no NW
$\omega_{\mathbf{k}}^{\pm1/2}$ nonlocality in the wave function.


\begin{thebibliography}{99}                                                                                               %


\bibitem {Lundeen}M. Tsang, Phys. Rev. Lett. \textbf{102}, 253601 (2009); J.
S. Lundeen, B. Sutherland, A. Patel, C. Stewart and C. Bamber, Nature
\textbf{474}, 188 (2011); S. Kocsis, B. Braverman, S. Ravets, M. J. Stevens,
R. P. Mirin, L. K. Shalm, and A. M. Steinberg, Science \textbf{332}, 1170 (2011).

\bibitem {BKP}D. Bohm, \textit{Quantum Theory} (Prentice-Hall, New Jersey,
1951); H. A. Kramers, \textit{Quantum Mechanics} (North-Holland, Amsterdam
1958); E. A. Power, \textit{Introductory Quantum Electrodynamics} (Longman,
London, 1964).

\bibitem {ScullyZubairy}M. O. Scully and M. S. Zubairy, \textit{Quantum
Optics} (Cambridge, U. K. 1996).

\bibitem {BB94}I. Bia\l ynicki-Birula, Acta Phys. Polonica A \textbf{86}, 97 (1994).

\bibitem {Sipe}J. E. Sipe, Phys. Rev A \textbf{52}, 1875 (1995).

\bibitem {BB96}I. Bia\l ynicki-Birula, \textit{Progress in Optics XXXVI,
}edited by E. Wolf (Elsevier,1996).

\bibitem {SmithRaymer}B. J. Smith and M. G. Raymer, New Journal of Physics
\textbf{9}, 414 (2007).

\bibitem {Gitman}S. P. Gavrilov and D. M. Gitman, Class. Quantum Grav.
\textbf{17} L133 (2000).

\bibitem {Mostafazadeh1}A. Mostafazadeh, Int. J. Mod. Phys. A21, 2553 (2006).

\bibitem {BirrellDavies}N. D. Birrell and P. C. W. Davies, \textit{Quanum
Fields in Curved Space} (Cambridge University Press, Cambridge, 1984).

\bibitem {Holland}P. R. Holland, \textit{The Quantum Theory of Motion}
(Cambridge University Press, 1993).

\bibitem {Colin}S. Colin, Bohm-Bell Beables for Quantum Field Theory, PhD
thesis (Vrije Universiteit Brussel, 2005).

\bibitem {Nikolic}H. Nikoli\'{c}, Found. Phys. \textbf{37}, 1563 (2007).

\bibitem {HalliwellOrtiz}J. Halliwell and M. Ortiz, Phys. Rev. D \textbf{48},
748 (1993).

\bibitem {HawtonPhysicaScripta}M. Hawton, Phys. Scr. \textbf{T147}, 014014 (2012).

\bibitem {Hegerfeldt}G.C.Hegerfeldt, Phys. Rev D 10, 3320-3321 (1974).

\bibitem {NewtonWigner}T. D. Newton and E. P. Wigner, Rev. Mod. Phys.
\textbf{21}, 400 (1949).

\bibitem {Pryce}M. H. L.Pryce, Proc. Roy. Soc. London, Ser. A \textbf{195}, 62 (1948).

\bibitem {Twisted}L. Allen, S.M.\ Barnett and M.J. Padgett, \textit{Optical
Angular Momentum} (Institute of Physics Bristol, 2003).

\bibitem {HawtonPosOp}M. Hawton, Phys. Rev. A, \textbf{59}, 954 (1999).

\bibitem {CohenQED1}C. Cohen-Tannoudji, J. Dupont-Roc and G. Grynberg,
\textit{Photons and Atoms: Introduction to Quantum Electrodynamics}
(Wiley-VCH, 1997).

\bibitem {HawtonBaylis}M. Hawton and W. E. Baylis, Phys. Rev. A, \textbf{64},
012101 (2001); \textbf{71}, 033816 (2005).

\bibitem {HawtonLorentzInv08}M. Hawton, Phys. Rev. A \textbf{78}, 012111 (2008).

\bibitem {Brody}D. C. Brody, J. Math. Phys. A: Math. Theor. \textbf{47},
035305 (2014).

\bibitem {quasi}Quasi-Hermitian operators are pseudo-Hermitian operators
restricted to a positive definite norm as required for the probability
interpretation of QM \cite{GeyerScholtz}.

\bibitem {GeyerScholtz}H.B. Geyer, W.D. Heiss, and F.G. Scholtz, Can. J. Phys.
\textbf{86}, 1195 (2008); A. Mostafazadeh, International Journal of Geometric
Methods in Modern Physics, \textbf{7}, 1191 (2010).

\bibitem {FeshbachVillars}H. Feshbach and F. Villars. Rev. Mod. Phys.
\textbf{30}, 24 (1958).

\bibitem {GGP}B. Gerlach, D. Gromes, and J. Petzold, Z. Phys. \textbf{204}, 1 (1967).

\bibitem {HenningWolf}J. J. Henning and W. Wolf, Z. Physik \textbf{242}, 12 (1971).

\bibitem {SUL}A. A. Semenov, C. V. Usenko, and B. I. Lev, Phy. Lett.
A\textbf{372}, 4180 (2008).

\bibitem {Mostafazadeh2}A. Mostafazadeh\textbf{ }and F. Zamani, Ann. Phys.
321, 2183 (2006).

\bibitem {PikeSarkar}E.R. Pike and S. Sarkar, \textit{The Quantum Theory of
Radiation (Oxford, 1995).}

\bibitem {ItzyksonZuber}C. Itzykson and J. B. Zuber, \textit{Quantum {Field}
Theory 1st ed. (McGraw-Hill, 1980).}

\bibitem {Fleming}G. N. Fleming, Phys. Rev. \textbf{137}, B188 (1965); Journ.
Math. Phys. \textbf{7,} 1959 (1966).

\bibitem {Glauber}R. J. Glauber, Phys. Rev. \textbf{130}, 2529 (1963).

\bibitem {Foldy}L. L. Foldy, Phys. Rev. \textbf{102}, 568 (1956).

\bibitem {Zamani}F. Zamani and A. Mostafazadeh, J. Math. Phys. \textbf{50},
052302 (2009).

\bibitem {LandauPeierls}L. D. Landau and R. Peierls, Z. Phys. \textbf{62}, 188
(1930); R.J. Cook, Phys. Rev. A \textbf{25}, 2164 (1982).

\bibitem {Sakurai}J. J. Sakurai, \textit{Modern Quantum Mechanics}
(Addison-Wesley, 1985).

\bibitem {NumDens95}M. Hawton and T. Melde, Phys. Rev. A \textbf{51}, 4186 (1995).

\bibitem {Debierre}V. Debierre, La Fonction D'Onde du Photon en Principe et en
Pratique, PhD thesis (\'{E}cole Centrale de Marseille, 2015).

\bibitem {DebierreDurt}V. Debierre and T. Durt, Phys. Rev. A \textbf{93},
023847 (2016).

\bibitem {Hawton07}M. Hawton, Phys. Rev. \textbf{A, 75}, 062107 (2007).
\end{thebibliography}
\end{document}